\documentstyle[twocolumn,prl,aps,epsf]{revtex} 

\begin{document}
\draft
\title{The first determination of Generalized Polarizabilities of \\ the proton by a Virtual Compton Scattering  experiment}

\author{
\setcounter{footnote}{0}
J.~Roche\protect\( ^{1}\protect \)\addtocounter{footnote}{+1}\footnotemark, 
 J.M.~Friedrich\protect\( ^{2}\protect \), 
D.~Lhuillier\protect\( ^{1}\protect \),
P.~Bartsch\protect\( ^{2}\protect \), D.~Baumann\protect\( ^{2}\protect \), J.~Berthot\protect\( ^{4}\protect \), 
P.Y.~Bertin\protect\( ^{4}\protect \), V.~Breton\protect\( ^{4}\protect \), 
W.U.~Boeglin\protect\( ^{2}\protect \)\footnotemark, R.~B\"ohm\protect\( ^{2}\protect \),  
N. D'Hose\protect\( ^{1}\protect \)\addtocounter{footnote}{-3}\footnotemark, S.~Derber\protect\( ^{2}\protect \), 
N.~Degrande\protect\( ^{3}\protect \),
  M.~Ding\protect\( ^{2}\protect \), M.O.~Distler\protect\( ^{2}\protect \), J.E.~Ducret\protect\( ^{1}\protect \),  
  I.~Ewald\protect\( ^{2}\protect \), H.~Fonvieille\protect\( ^{4}\protect \), J.~Friedrich\protect\( ^{2}\protect \),
 P.A.M.~Guichon\protect\( ^{1}\protect \), H.~Holvoet\protect\( ^{3}\protect \), Ch.E.~Hyde-Wright\protect\( ^{5}\protect \),
 P.~Jennewein\protect\( ^{2}\protect \), M.~Kahrau\protect\( ^{2}\protect \), S.~Kerhoas\protect\( ^{1}\protect \),  
 K.W.~Krygier\protect\( ^{2}\protect \),
  B.~Lannoy\protect\( ^{3}\protect \),  A.~Liesenfeld\protect\( ^{2}\protect \), C.~Marchand\protect\( ^{1}\protect \), 
D.~Marchand\protect\( ^{1}\protect \)\addtocounter{footnote}{+2}\footnotemark, 
J.~Marroncle\protect\( ^{1}\protect \), J.~Martino\protect\( ^{1}\protect \),
 H.~Merkel\protect\( ^{2}\protect \),  P.~Merle\protect\( ^{2}\protect \), G.~De~Meyer\protect\( ^{3}\protect \), 
 J.~Mougey\protect\( ^{1}\protect \)\footnotemark, U.~M\"uller\protect\( ^{2}\protect \), 
 R.~Neuhausen\protect\( ^{2}\protect \),
  Th.~Pospischil\protect\( ^{2}\protect \), G.~Quemener\protect\( ^{4}\protect \)\addtocounter{footnote}{-1}\footnotemark, 
  O.~Ravel\protect\( ^{4}\protect \)\footnotemark, Y.~Roblin\protect\( ^{4}\protect \),  D.~Rohe\protect\( ^{2}\protect \),
G.~Rosner\protect\( ^{2}\protect \)\footnotemark, D.~Ryckbosch\protect\( ^{3}\protect \), 
  H.~Schmieden\protect\( ^{2}\protect \),  G.~Tamas\protect\( ^{2}\protect \), M.~Tytgat\protect\( ^{3}\protect \), 
  M.~Vanderhaeghen\protect\( ^{2}\protect \), 
 L.~Van Hoorebeke\protect\( ^{3}\protect \), R.~Van de Vyver\protect\( ^{3}\protect \), 
 J.~Van de Wiele\protect\( ^{6}\protect \), P.~Vernin\protect\( ^{1}\protect \),
 A.~Wagner\protect\( ^{2}\protect \), Th.~Walcher\protect\( ^{2}\protect \) and M.~Weiss\protect\( ^{2}\protect \) }

\address{{ \protect\( ^{1}\protect \)CEA Saclay, DSM/DAPNIA/SPhN, 91191 Gif-sur-Yvette Cedex, France} }
\address{{ \protect\( ^{2}\protect \)Institut f\"ur Kernphysik, Universit\"at Mainz, 55099 Mainz, Germany} }
\address{{ \protect\( ^{3}\protect \)FWO and SSF, University of Gent, Proeftuinstraat 86, 9000 Gent, Belgium} }
\address{{ \protect\( ^{4}\protect \)LPC de Clermont-Fd, IN2P3-CNRS, Universit\'e Blaise Pascal, 63177 Aubiere Cedex, France} }
\address{{ \protect\( ^{5}\protect \)Old Dominion University, Norfolk,  VA23529, U.S.A.} }
\address{{ \protect\( ^{6}\protect \)IPN d'Orsay, IN2P3-CNRS, Universit\'e de Paris Sud, 91406 Orsay Cedex, France} }

\maketitle

\begin{abstract}
Absolute differential cross sections for the reaction $e p \rightarrow e p \gamma$ have been measured at a  four-momentum 
transfer with virtuality  $Q^2=0.33$ GeV$^2$ and polarization $\epsilon = 0.62$ in the range 33.6 to 111.5 MeV/c for the 
momentum of the outgoing photon in the photon-proton center of mass frame.
The experiment has been performed with the high resolution spectrometers at the Mainz Microtron MAMI. From the photon 
angular   distributions, two structure functions which are a linear combination of the generalized polarizabilities have 
been determined for the first time.

PACS numbers: 13.60.Fz, 14.20.Dh, 13.10.+q, 25.30.Rw
\end{abstract}


Polarizabilities are fundamental quantities that characterize the response of a composite system to static or slowly varying
 external electric or magnetic fields. For the proton, Compton scattering of real photons has been used to determine the 
 electric and magnetic polarizabilities~\cite{rcs}.  In this letter, we report measurements of Compton scattering of virtual
  photons, leading to the first determination of generalized polarizabilities (GPs).
Virtual Compton scattering (VCS) off the proton refers to the reaction $\gamma ^* p \rightarrow p \gamma$ where $\gamma ^*$ 
stands for an incoming virtual photon of four-momentum squared $Q^2$. This reaction is experimentally accessed through  
photon electroproduction  $e p \rightarrow e p \gamma$. 
We study this reaction in the regime where the produced photon has a small enough energy that its electric (E) and magnetic 
(M) fields look constant over the size of the nucleon. In this regime the reaction can be interpreted as electron scattering
 on a nucleon placed in a quasi-constant applied EM field~\cite{pammarc}. The induced motion of the nucleon as a whole  can 
 be eliminated thanks to a low energy theorem~\cite{low}, so one is left with the deformation, due to the applied field, of 
 the nucleon internal currents $\delta J^{\mu}(r)$ and the electron scattering measures its Fourier transform $\delta 
 J^{\mu}(Q)$. To lowest order in $\alpha_{QED}$, 
$\delta J^{\mu}(Q)$ is linear in the applied field and the 6 coefficents of proportionality are the GPs~\cite{pammarc,pam,dre}. 
When $Q^2$=0 two of them reduce to the usual polarizabilities 
$\alpha_E$ and $\beta_M$ measured in real Compton scattering~\cite{rcs}.
A measurement of how the nucleon's internal current distribution is deformed by an external EM field will yield valuable 
information about its non perturbative structure. This is illustrated by the quite different results obtained for the GPs 
according to the used model~\cite{hem97,met96,vdh96,liu96,barbara}. As an example of new information provided by VCS and 
which could not be obtained either from ordinary electron scattering or real Compton scattering, consider the case of 
Coulomb scattering  on a nucleon placed in a magnetic field. The action of the latter is mainly to flip the spin of the 
quarks, which hardly modifies the charge density. Therefore  the corresponding GPs (namely
${P}^{(11,00)1}$ and ${P}^{(11,02)1}$ in Eq.~\ref{combi_pol}) receive no contribution from the quarks. This is confirmed 
by actual calculations. As a consequence the value of these GPs should be mainly determined by the pion cloud.

First indications of the feasibility of a VCS experiment were reported in \cite{vandenbrand}. However only recently, with 
the advent of high-intensity continuous beam electron accelerators, a thorough experimental investigation of the VCS process
 became possible. This letter reports on the first VCS experiment dedicated to the GPs which has been  performed at the Mainz 
 Microtron MAMI.

The general theoretical framework for VCS has been extensively described by Guichon {\em et al.} \cite{pam,pammarc} and  
by Drechsel {\em et al.} \cite{dre} and only the relevant issues will be discussed here. 
In the reaction $e p \rightarrow e p \gamma$, the final photon can be emitted either by the electron, which is called   
the Bethe-Heitler (BH) process, or by the proton. 
The latter process is the VCS process which consists of Born  and Non-Born amplitudes; while the Born amplitude  depends 
only on  static properties of the proton (charge, mass,  form factors),  the Non-Born amplitude contains dynamical  internal
 structure information in terms of GPs.
Despite of the strong dominance of the BH process, the VCS process can be measured under appropriately chosen kinematics 
through the interference with the dominant BH amplitude. 
In the  zero-energy limit of the final photon, the cross section is independent of the dynamical nucleon structure~\cite{low},
 and  can be evaluated using only the known  BH and Born amplitudes.   This is summarized by the following expression:
\begin{eqnarray}
d^5\sigma^{exp} (\mathrm q, \mathrm q', \epsilon, \theta, \varphi) & = d^5\sigma^{BH+Born} (\mathrm q, \mathrm q', 
\epsilon, \theta, \varphi) \nonumber \\
 & + \phi \mathrm q'  \Psi_0(\mathrm q,\epsilon,\theta,\varphi) +  \mathcal{O}(\mathrm q'^2)
\label{eq1}
\end{eqnarray}
where $d^5\sigma$ is a notation for the differential cross  section  ${d^5\sigma}/{d \mathrm k'_{lab} [d\Omega_e]_{lab} 
[d\Omega_p]_{CM}}$
wherein $\mathrm k'_{lab}$ is the absolute value of the outgoing electron momentum in the laboratory,
 $\mathrm q$ and $\mathrm q'$ are the absolute values of the three-momenta of the virtual and real photons (in the 
 photon-proton center of mass (CM) system) respectively and $\epsilon$ is the virtual photon polarization.
$\theta$ is the angle between the real and virtual photon in the CM
 system  and  $\varphi$ is the angle between the electron  and the photon-proton plane.  $\phi$ stands for a phase space 
 factor. $\Psi_0(\mathrm q,\epsilon,\theta,\varphi)$  is the leading term in the expansion in powers of the real photon 
 momentum $\mathrm q'$. It  contains the dynamical internal structure information of the proton, expressed  by 6 GPs. 
 The latter  are denoted by $P^{(\rho'L', \rho L)S}(\mathrm q)$ where $L (L')$ are the initial (final) 
 photon orbital angular momentum, $\rho (\rho')$ the type of multipole transition (0 for Coulomb, 1 for magnetic) , and $S$ 
 distinguishes between non-spin-flip ($S=0$)  and spin-flip ($S=1$)  transitions at the nucleon side. One has  2 
 non-spin-flip GPs, 
${P}^{(01,01)0}$, 
${P}^{(11,11)0}$, proportional  to $\alpha_E$ and $\beta_M$ at $Q^2=0$ respectively,
and 4 spin-flip GPs, 
${P}^{(11,11)1}$,
${P}^{(11,00)1}$,
${P}^{(11,02)1}$,
${P}^{(01,12)1}$.
The GPs are functions of $\mathrm q$ (or   $\tilde Q^2 = Q^2|_{\mathrm q'=0}$).
In an unpolarized measurement, $\Psi_0(\mathrm q,\epsilon,\theta,\varphi)$ can be written as:
\begin{eqnarray}
\Psi_0(\mathrm q,\epsilon,\theta,\varphi) & = v_1(\theta,\varphi,\epsilon) (P_{LL}(\mathrm q)-P_{TT}(\mathrm q)/\epsilon) 
\nonumber \\
& + v_2(\theta,\varphi,\epsilon) P_{LT}(\mathrm q)
\label{eq2}
\end{eqnarray}
where $v_1(\theta,\varphi,\epsilon)$, $v_2(\theta,\varphi,\epsilon)$ are known kinematical factors and $P_{LL}(\mathrm q)$, 
$P_{TT}(\mathrm q)$ and $P_{LT}(\mathrm q)$ are structure functions defined by:
\begin{eqnarray}
P_{LL}= & - 2\sqrt{6}m G_E {P}^{(01,01)0} \nonumber \\   
P_{TT}= &  {3} G_M {\mathbf {\mbox q}}^2 
\left( \sqrt{2} {P}^{(01,12)1} 
-  {P}^{(11,11)1} /{\tilde{q}_0}  \right) \nonumber \\ 
P_{LT}= & \sqrt{\frac{3}{2}}  \frac{m \mathbf {\mbox q}}{\tilde {Q}} G_E {P}^{(11,11)0} \nonumber \\
 & + \frac{\sqrt{3}}{2} \frac{\tilde {Q}}{ \mathbf {\mbox q}}   G_M ( {P}^{(11,00)1}  + \frac{\mathrm{q^2}}{\sqrt{2}} 
 {P}^{(11,02)1} )    
\label{combi_pol}
 \end{eqnarray}
where $m$ stands for the proton mass, $G_E$ and $G_M$ denote the form factors  evaluated at $\tilde Q^2$ and $\tilde {q}_0$ is 
the CM virtual photon energy at $\mathrm q'=0$.

Absolute cross sections $d^5\sigma^{exp}$~\cite{david,julie,jan} have been measured using the three-spectrometer 
facility~\cite{blom} of the A1 collaboration at the 855 MeV Mainz Microtron MAMI.
The scattered electron and the recoiling proton were detected in coincidence with  two of the high-resolution magnetic 
spectrometers. The photon production process was selected by a cut on the missing mass around zero, which was possible 
thanks to the excellent resolution of the facility (momentum resolution  of $10^{-4}$ and  angular resolution better than 
3 mrad). Liquid hydrogen served as a target, contained in a target cell of  49.5 mm total length  with a havar wall 
thickness of  9 $\mu$m. The use of typical electron currents of 30 $\mu$A yielded a luminosity of ${\cal L}=4 \cdot 10^{37}$
 cm$^{-2}$s$^{-1}$.

In this first VCS experiment below pion threshold,  the five-fold differential cross section was  measured in a wide  
angular range at 5 values of the photon momentum $\mathrm q'$: 33.6, 45.0, 67.5, 90.0, and 111.5 MeV/c. Two kinematical 
variables were kept fixed, namely the virtual photon momentum, q = 600 MeV/c ($\tilde Q ^2$ = 0.33 GeV$^2$) and  the 
polarization $\epsilon$ = 0.62.  The out-of-plane angle $\varphi$ range is determined by the acceptance of the  two 
spectrometers around 0$^\circ$ and 180$^\circ$.  To ease the presentation (see Fig. 1), the data are plotted with 
$\theta$ ranging from -180$^\circ$ to +180$^\circ$, the negative values correspond to  $\varphi = 180^{\circ}$.  
The wide range of $\theta$ from -141$^\circ$ to +6$^\circ$  was covered by only  changing the setting of the proton 
spectrometer. This angular range covers the backward direction relative to the incoming and outgoing electron. Here 
the VCS contributions are dominant because the BH radiation
is mainly emitted  in the electrons directions. 

The influence of the GPs gives rise to  
 a small deviation of the measured cross section from  the BH+Born cross section.   Theoretically it is expected to be  
 about 10\%  at the highest $\mathrm q'$ value. The cross sections at each bin are determined within a statistical accuracy 
 of 3\%; however, a careful analysis of possible systematic errors on the above  deviation is of particular importance. 
(1) The proton form factors are not  exactly known. Consequently, we also measured the absolute elastic scattering cross 
section for each   kinematical setting of the VCS experiment. These measurements validate 
the use of the form factor parametrization from H\"ohler~\cite{hoehler} at a precision better than $\pm$ 1\%.
(2) The luminosity and the detector efficiencies are controlled within  the same accuracy.
(3) The radiative corrections, which are of the order of  20\% of the cross section, have been calculated  by Vanderhaeghen 
{\it et al.}~\cite{marc_rad} taking into account all the diagrams up to  order $\alpha^4$ in the VCS cross section. The 
corresponding systematic uncertainties are estimated to equal  $\pm 2\%$.
(4) An extensive Monte Carlo simulation~\cite{luc} has been performed in order to verify the missing mass spectra and to 
determine the solid angles within  an accuracy of $\pm$2\%.
The code generates events according to  the BH+Born cross section   and takes into account all resolution deteriorating 
effects and real photon radiations.
(5) While the  uncertainties (1)-(4) are constant over the angular range of the real photon,  small imperfections in the 
spectrometer optics calibration produce distortions of the angular distributions: these effects lead to an estimated change 
in absolute cross section of $\pm$2.5\%.

Fig. 1 shows the measured differential cross sections (with statistical errors only)   as a function of $\theta$ for 5 
values of the real photon  momentum.  The  cross sections $d^5\sigma^{BH+Born}$, calculated using the H\"ohler 
form factors,    are presented by the solid lines. 
At the smallest photon momentum, $\mathrm q'$ = 33.6 MeV/c, the agreement between the data and $d^5\sigma^{BH+Born}$ is 
excellent,
while with increasing $\mathrm q'$ one observes growing deviations from this known cross section. This represents the 
effect of the polarizabilities.

\begin{figure}[htb]
\vglue -.5cm
\epsfxsize=9cm\epsfbox{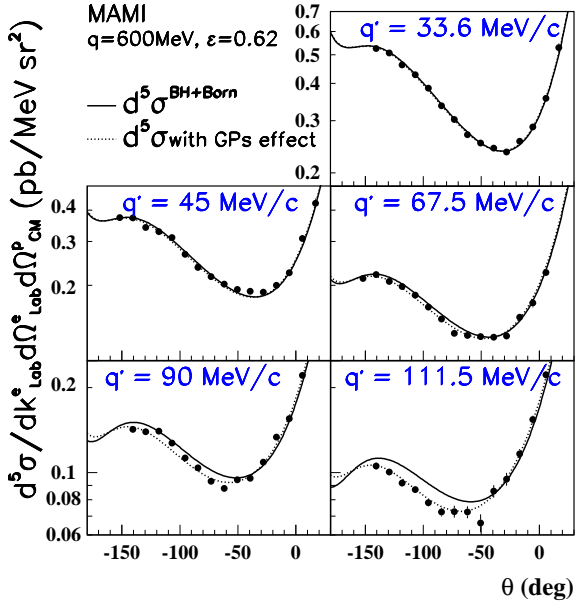}
\caption {   Differential cross section for the reaction $e p \rightarrow e p \gamma$  measured at MAMI as a function of 
$\theta$ for $\mathrm q$ = 600 MeV/c or $\tilde Q^2$ = 0.33 GeV$^2$, $\epsilon$ =0.62  and for five values of the real 
photon momentum $\mathrm q'$. The known  cross section $d^5\sigma^{BH+Born}$ is indicated by the solid lines. The 
experimental data points  deviate from the solid lines as $\mathrm q'$ increases; this is the effect of the proton 
polarizabilities.
The dotted lines represent the theoretical cross sections including the effect of the GPS, determined in  this experiment.}
\label{fig1}
\end{figure}

Fig. 2 shows  $(d^5\sigma - d^5\sigma^{BH+Born})/\phi \mathrm q'$ as a function of the real photon momentum $\mathrm q'$ at 
the 14 measured   angles $\theta$, of which the intercept with the ordinate axis directly yields $\Psi_0$ in Eq. (1). 
Several methods were applied in order to determine the intercept, which  will be discussed in a forthcoming paper.
The methods that were tried, taking into account a possible $\mathrm q'$ dependence of 
$(d^5\sigma - d^5\sigma^{BH+Born})/\phi \mathrm q'$, all show that this dependence is weak. As such, here  the most simple 
approach was used, i.e.  no $\mathrm q'$ dependence.
$\Psi_0$ is then determined at each   angle $\theta$  by the weighted mean value of the data at the 5 photon energies.  
Fig. 3 presents the resulting values of $\Psi_0/v_2$ as a function of $v_1/v_2$ (cf. Eq. (2)). The data are well aligned,  
this allows us  to extract 
the two structure functions $P_{LL}-P_{TT}/\epsilon$ and $P_{LT}$ as the slope and intercept, respectively,  of a linear fit 
to the data. 

Table 1 shows the results with a statistical error and two systematic errors, the first one corresponds  to the 
 normalisation of the  angular distributions,  the second one stems from the 
 distortion  of the distributions. 
 All these effects will be  evaluated in detail in a forthcoming paper. 
The other methods for obtaining the combinations give within
the error bars compatible results with the ones that are presented here. 
The dotted lines in Fig.~1 are the theoretical 
 cross sections calculated using the  two structure functions measured in  this experiment.

\begin{figure}[htb]
\vglue -.5cm
\hglue .3cm
\epsfxsize=9cm\epsfbox{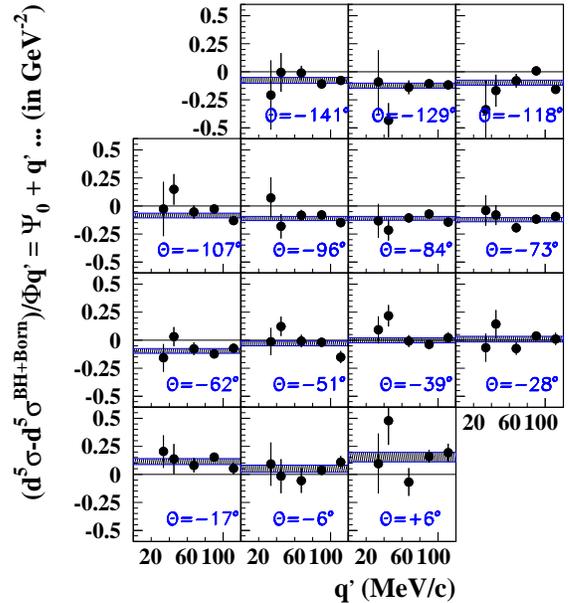}
\caption {   $(d^5\sigma - d^5\sigma^{BH+Born})/\phi \mathrm q'$ measured as  function of the real photon momentum 
$\mathrm q'$ at 14 angles $\theta$. The intercept with the  ordinate is $\Psi_0$. 
The shaded bands represent the uncertainty in the extrapolation to $\mathrm q'$ = 0 by a constant.
}
\label{fig2}
\end{figure}

Our results for the two structure functions are also compared in Table 1 with different model predictions: a heavy-baryon 
chiral perturbation theory calculation (HBChPT)~\cite{hem97},  the linear sigma
model (LSM)~\cite{met96}, an effective lagrangian model (ELM)~\cite{vdh96} and  two non relativistic constituent quark 
models (NRCQM)~\cite{liu96,barbara}). 
The results seem to favour the HBChPT  which at least for $P_{LT}$ may not be so surprising since this approach is based on  
chiral symmetry and therefore should correctly  describe the pion cloud which dominates 
$P^{(11,00)1}$ and $P^{(11,02)1}$, as it was pointed out in the introduction. 
Note that the LSM model also respects   chiral symmetry but it contains otherwise too restrictive hypotheses.  The apparent 
agreement of the NRCQM models for $P_{LT}$ is accidental. It is known that at $Q^2=0$  these models overshoot 
the experimental value of $P^{(11,11)0}$ by at least a factor 2.

To summarize, results are reported on the first VCS experiment which   allows the determination of two dynamical structure 
functions which are a linear combination of the GPs. The measurement has been performed at MAMI at 
$Q^2$ = 0.33 GeV$^2$ and gives results compatible with chiral perturbation predictions.   
Two other  experiments, aiming at  the determination of the same structure functions are performed at different $Q^2$: the 
Jefferson Lab. experiment~\cite{exp_cebaf} at $Q^2$ = 1 and 2 GeV$^2$ and the MIT-Bates experiment~\cite{exp_mit} at $Q^2$ 
= 0.05 GeV$^2$. However, to  measure independently  the 6 GPs and separate non-spin-flip and spin-flip 
contributions, it will be necessary to perform double polarization experiments.

\begin{figure}[htb]
\vglue -.5cm
\hglue 0cm
\epsfxsize=8.0cm\epsfbox{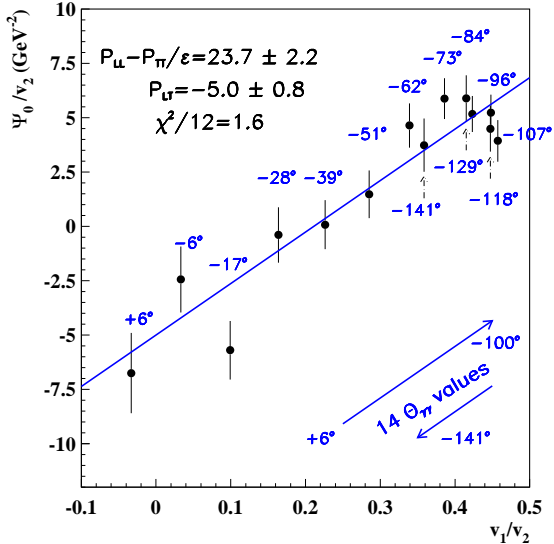}
\caption{    $\Psi_0/v2$ as a function of $v1/v2$.
The  errors  are statistical only. 
The line represents the linear fit to the data according Eq. (2) with the values for $P_{LL}-P_{TT}/\epsilon$ and $P_{LT}$ 
shown in the graph.
}
\label{fig3}
\end{figure}

The authors wish to acknowledge the excellent support of the accelerator group of MAMI. 
This work was supported in part by the French CEA and CNRS/IN2P3, the Deutsche Forschungsgemeinschaft (SFB 201), 
the FWO-Flanders (Belgium),  the BOF-Gent University, 
the European Commission ERB FMRX-CT96-0008, the US DOE
and the US NSF.

\newpage

\begin{table}
\caption{The structure functions determined in this experiment and compared to model predictions at $Q^2$ = 0.33 GeV$^2$ 
and $\epsilon=0.62$. The first error is statistical only, while the two others are systematic as commented in the text. 
Previous results[1] at $Q^2=0$ and comparison to the  same  models are also presented.
}
\label{tab1}
\begin{tabular}{ccc}
        &  $P_{LL}(Q^2)-\frac{1}{\varepsilon} P_{TT}(Q^2)$ & $ P_{LT}(Q^2) $ \\ 
        &  (in  GeV$^{-2}$)                                   & (in GeV$^{-2}$)    \\ 
\hline
& $Q^2=0.33$ GeV$^2$ & \\
\hline
This experiment     & 23.7                           &   -5.0   \\
                    &  $\pm$ 2.2 $\pm$ 0.6 $\pm$ 4.3 &     $\pm$ 0.8 $\pm$ 1.1$\pm$ 1.4 \\  
HBChPT~\cite{hem97} & 26.0           & -5.3 \\
LSM~\cite{met96}    & 11.5           & ~0.0 \\ 
ELM~\cite{vdh96}    & 5.9            & -1.9 \\ 
NRCQM~\cite{liu96}  & 11.1           & -3.5 \\
NRCQM~\cite{barbara}  & 14.9           & -4.5 \\
\hline
& $Q^2=0 $ GeV$^2$ &  \\
\hline
Old experiments~\cite{rcs}     & 81.0                           &   -7.0   \\
                    &  $\pm$ 5.4 $\pm$ 3.3  &     $\mp$ 2.7 $\mp$ 1.7 \\  
HBChPT~\cite{hem97} & 83.5           & -4.2 \\
LSM~\cite{met96}    & 49.8           & +6.6 \\ 
ELM~\cite{vdh96}    & 47.8            & -5.2 \\ 
NRCQM~\cite{liu96}  & 37.1           & -13.1 \\
NRCQM~\cite{barbara}  & 37.0           & -15.8 \\
\end{tabular}
\end{table}

\end{document}